\documentclass[final]{IEEEtran}
\usepackage{cite}
\usepackage{color}
\usepackage{threeparttable}
\usepackage{graphicx}
\usepackage{picinpar}
\usepackage[cmex10]{amsmath}
\usepackage{amsmath,amsfonts,amssymb}
\usepackage{subfigure}
\usepackage{changepage}
\usepackage{algorithm}
\usepackage{algorithmic}
\usepackage{stfloats}
\usepackage{bm}
\usepackage{enumerate}

\begin{document}
\newtheorem{lemma}{Lemma}
\newtheorem{corol}{Corollary}
\newtheorem{theorem}{Theorem}
\newtheorem{proposition}{Proposition}
\newtheorem{definition}{Definition}
\newcommand{\e}{\begin{equation}}
\newcommand{\ee}{\end{equation}}
\newcommand{\eqn}{\begin{eqnarray}}
\newcommand{\eeqn}{\end{eqnarray}}

\newenvironment{shrinkeq}[1]
{ \bgroup
\addtolength\abovedisplayshortskip{#1}
\addtolength\abovedisplayskip{#1}
\addtolength\belowdisplayshortskip{#1}
\addtolength\belowdisplayskip{#1}}
{\egroup\ignorespacesafterend}

\title{GMD-Based Hybrid Beamforming for Large Reconfigurable Intelligent  Surface Assisted Millimeter-Wave Massive MIMO }

\author{Keke Ying, Zhen Gao,~\IEEEmembership{Member,~IEEE}, Shanxiang Lyu,~\IEEEmembership{Member,~IEEE}, Yongpeng Wu,~\IEEEmembership{Senior Member,~IEEE}, Hua Wang,~\IEEEmembership{Member,~IEEE}, and Mohamed-Slim Alouini,~\IEEEmembership{Fellow,~IEEE}
	\thanks{K. Ying and Z. Gao are with both Advanced Research Institute of Multidisciplinary Science (ARIMS) and School of Information and Electronics,
		Beijing Institute of Technology (BIT), Beijing 100081, China (E-mail: 1120161199bit.edu.cn, gaozhen16@bit.edu.cn).}
	\thanks{S. Lyu is with the College of Cyber Security, Jinan University, Guangzhou 510632, China, and also with the State Key Laboratory of Cryptology, P.O.Box 5159, Beijing, 100878, China.}
	\thanks{Y. Wu is with Department of Electronic Engineering, Shanghai Jiao Tong University, Shanghai 200240, China.}
	\thanks{H. Wang is with the School of Information and Electronics, Beijing Institute of Technology, Beijing 100081, China. }	
	\thanks{M.-S. Alouini is with the Electrical Engineering Program, Division of Physical Sciences and Engineering, King Abdullah University of Science and Technology, Thuwal 23955, Saudi Arabia.}
}

\maketitle
\begin{abstract}
Reconfigurable intelligent surface (RIS) is considered to be an energy-efficient approach to reshape the wireless environment for improved throughput. Its passive feature greatly reduces the energy consumption, which makes RIS a promising technique for enabling the future smart city. Existing beamforming designs for RIS mainly focus on optimizing the spectral efficiency for single carrier systems. To avoid the complicated bit allocation on different spatial domain subchannels in MIMO systems, in this paper, we propose a geometric mean decomposition-based beamforming for RIS-assisted millimeter wave (mmWave) hybrid MIMO systems so that multiple parallel data streams in the spatial domain can be considered to have the same channel gain. Moreover, by exploiting the common angular-domain sparsity of mmWave massive MIMO channels over different subcarriers, a simultaneous orthogonal match pursuit algorithm is utilized to obtain the optimal multiple beams from an oversampling 2D-DFT codebook. Besides, by only leveraging the angle of arrival and angle of departure associated with the line of sight (LoS) channels, we further design the phase shifters for RIS by maximizing the array gain for LoS channel. Simulation results show that the  proposed scheme can achieve better BER performance than conventional approaches. Our work is an initial attempt to discuss the broadband hybrid beamforming for RIS-assisted mmWave hybrid MIMO systems. 
%Our work is an initial attempt to discuss the broadband hybrid beamforming for RIS-assisted mmWave hybrid MIMO systems.%
\end{abstract}


\IEEEpeerreviewmaketitle

\section{Introduction}
\label{sec:introduction}
Reconfigurable intelligent surface (RIS) has appeared to be an effective technology to improve the capacity and channel condition in the future wireless communications \cite{summary}. Its passive and low-cost characteristics make it be employed in a wide plethora of applications including millimeter-wave (mmWave) cellular networks \cite{liao}, Internet of Things \cite{gao_cs}, even green mobile edge computing \cite{add1}. RIS is composed of a large number of   passive reflecting units, and by changing the reflection phase and amplitude of incident signals, these units can improve the transmission performance of wireless system in an intelligent approach.

Existing beamforming designs for RIS aim at designing the reflecting matrix at the RIS or jointly designing the active and passive beamformers at the base station (BS) and RIS \cite{{irs_beamforming},{irs1},{irs2},{irs3},{irs4}}. Most of these beamforming problems are formulated as optimization problems, where a specific target, such as spectral efficiency (SE) \cite{{irs_beamforming},{irs2}}, power consumption \cite{irs1}, or receiver signal-to-noise-ratio (SNR) \cite{{irs3},{irs4}} is considered to optimize. Therefore, convex optimization or some iterative algorithms for non-convex targets can be used.
However, most works are designed under a narrowband channel with fully-digital structure multiple input multiple output (MIMO), where the hardware cost is prohibitively high \cite{ke}, while a wideband channel is seldom discussed. For wideband channels, an orthogonal frequency division multiplexing (OFDM) based channel estimation and passive beamforming scheme has been designed in \cite{irs_ofdm}, where the semidefinite relaxation (SDR) is used to optimize the upper bound of achievable rate. Nevertheless, the beamformer at the BS has not been considered in this occasion.
To the best of our knowledge, these aforementioned beamforming for RIS scenarios are considered in the low-frequency bands, and the mmWave MIMO with hybrid analog-digital structure \cite{{wan}} has not been considered.

In mmWave MIMO beamforming\footnote{For convenience, in this paper, we use ``beamforming" to unify the design of the beamformer at the transmitter and combiner at the receiver. Furthermore, in order to distinguish baseband and RF parts, we use ``baseband/digital beamformer/combiner" and ``RF/analog beamformer/combiner" respectively in hybrid MIMO architectures. Also see the description of transmission model in Section \ref{sec:system model}.} without RIS, the main target is to eliminate the interference among data streams or users for improved throughput. Hybrid beamforming on narrowband channel has been well investigated in \cite{{gmd},{sparse},{somp},{codebook}}.  In \cite{sparse}, an orthogonal match pursuit (OMP) algorithm based hybrid beamforming is applied to achieve the performance close to the optimal full-digital beamforming. In \cite{somp}, a codebook based beamspace singular value decomposition (SVD) and hybrid beamforming are proposed to avoid the prohibitive matrix inversion computation. Different from \cite{sparse} and \cite{somp}, \cite{gmd} and \cite{codebook} aim to optimize the bit error rate (BER) performance, where a geometric mean decomposition (GMD)-based baseband beamforming and codebook-based analog beamforming is respectively adopted. In view of the frequency-selective-fading channels in practice, OFDM is used to combat the multipath effect in the wideband systems. The authors in \cite{heath1} prove that semi-unitary frequency flat beamforming and combining are sufficient to achieve maximum SE when there are not too much scatterers in the frequency-selective channel. \cite{heath2} proposes a limited feedback for channel state information (CSI) and presents a hybrid analog-digital beamforming codebook design for wideband case.
Some other solutions to further reduce the cost of hardware or feedback, such as dynamic partially-connected structure or low-resolution hybrid beamforming have also been considered in \cite{{pca},{heath4},{dai}}.

Now with the emergence of RIS, beamforming becomes more flexible and controllable. This gives us the inspiration to combine the typical hybrid beamforming with the passive RIS beamforming together. In this situation,  both the advantage of large bandwidth in mmWave and the blockage effect improvement benefited from RIS can be utilized, which can lead to a significant improvement in throughput and coverage.

Since the relative position between the BS and RIS has been determined physically, the channel condition between them remains almost unchanged during the communication. Therefore, the main uncertainty results from the links between the RIS and user equipments (UEs). However, RIS is usually assumed to be a large-scale array, which makes the pilot overhead in channel estimation extremely high. Due to the high hardware cost and pilot overhead in channel estimation \cite{{estimation1},{heath3},{fei}}, a reflecting matrix design without perfect CSI or with partial channel information needs to be designed.
Besides, most prior works about passive beamforming consider a fully-digital MIMO structure at the BS, and the UE is assumed to be equipped with a single antenna.
Combined with the well-studied hybrid beamforming for conventional MIMO system without RIS, a more generalized scenario is necessary to be discussed.

In this paper, a hybrid MIMO-OFDM system working at mmWave frequency with RIS is considered. The RIS reflecting coefficient matrix and hybrid beamformer/combiner at both the BS and UE are separately designed. To the best of our knowledge, this is the first paper to investigate the beamforming for RIS-assisted hybrid MIMO systems. To be more specific, for the design of digital baseband beamformer/combiner, we consider a GMD approach, which has been proved to be an effective way to avoid the complicated bit/power allocation and can achieve a better BER performance than SVD \cite{gmd}. For the analog part, we adopt a simultaneous orthogonal match pursuit (SOMP) algorithm \cite{greedy}, which is an extension of the classical OMP algorithm \cite{sparse}, to choose multiple optimal beams from a pre-defined codebook. Note that  since both the BS and UE adopt the uniform planar array (UPA) in this work, we adopt an oversampling 2D-discrete Fourier transform (DFT) codebook, which can effectively avoid the impractical priori information of all the steering vectors of the MIMO channels required by conventional compressive sensing (CS)-based hybrid beamforming designs \cite{{sparse},{gmd}}. In addition,
the angle parameters of non-line-of-sight (NLoS) paths are difficult to acquire in practice. To reduce the computational complexity of RIS reflection matrix, we will only leverage the angle of arrival (AoA) of the line-of-sight (LoS) BS-RIS channel and the and the angle of departure (AoD) of the LoS RIS-UE channel to design the reflecting matrix.  Simulation results show that the proposed scheme can achieve a satisfactory performance in the wideband RIS scenarios, which makes it possible for RIS to be widely used in the future smart mobile communications.

The rest of this paper is organized as follows. The system model is briefly introduced in Section II. Section III presents the hybrid beamforming scheme at BS/UE and passive beamforming design at RIS.
Section IV evaluates the performance of proposed scheme through simulations and comparison. Finally, we conclude this paper in Section V.

\textit{Notations}: Lower-case and upper-case boldface letters denote vectors and matrices, respectively; ${\left(  \cdot  \right)\!^T}$, ${\left( \cdot \right)\!^*}$, ${\left( \cdot \right)\!^{\dagger}}$, and ${\left( \cdot \right)\!^H}$ denote the transpose, conjugate, pseudo-inverse, and conjugate transpose of a matrix, respectively; ${\left(  \cdot  \right)^{(i)}}$ and ${\left(  \cdot  \right)\!_{i,j}}$ represent the ${i}$-th column and ${i}$-th row and ${j}$-th column element of a matrix, respectively; ${\mathbf{I}}_N$ represents the ${N \!\times \!N}$ identity matrix; ${\left\| {\cdot} \right\|_{\mathrm F}}$ and diag$ \left(  \cdot  \right)$ represent Frobenius norm and diagonalization, respectively.

\section{System Model}
\label{sec:system model}
\begin{figure}
	\centering
	\includegraphics[width=8cm, keepaspectratio]%
	{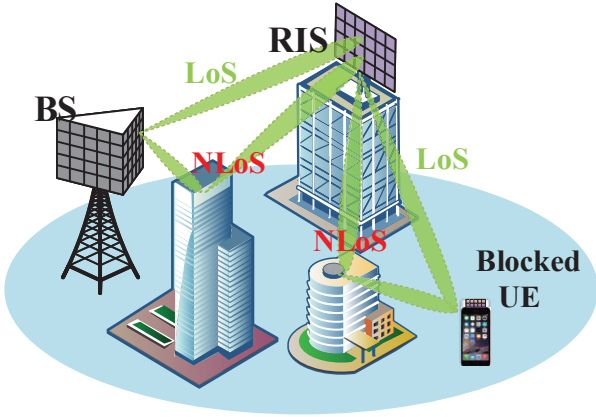}
	\vspace*{-1.0mm}
	\caption{A diagram of mmWave MIMO system with the assistance of RIS for a blocked UE.}
	\label{fig1}
	\vspace*{-3.0mm}
\end{figure}

We consider a single-user mmWave hybrid MIMO system adopting OFDM to combat the time dispersive effect over wideband channels. As shown in Fig.\ref{fig1}, an RIS is deployed between the BS and UE. The LoS path between the BS and UE is blocked by some buildings. Therefore, the UE will receive signals from the BS via RIS.
We assume that the BS and UE are both equipped
with the UPAs. The BS is equipped with ${N_t}$ antennas but ${M_t} \ll {N_t}$ radio frequency (RF) chains, and the UE employs ${N_r}$ antennas but ${M_r} \ll {N_r}$ RF chains to support ${N_s}\le{M_r}$ data streams. Physically, each RF chain is connected to $ N_t $ $(N_r)$ antennas through $N_t$ $(N_r)$ phase shifters at the BS (UE).
The RIS is assumed to be comprised of ${N_u}$ passive reflecting elements, and each of which can induce a phase shift to the incident signal independently with the help of an RIS controller.
%Thus, we can regulate the multipath channel environment actively instead of enduring the scattering surroundings passively.

The mmWave MIMO channel is assumed to be a sum of the contributions of one LoS path  and ${N_c}$ NLoS scattering clusters. Each scattering cluster contains ${N_p}$ propagation paths with a corresponding relative time delay. Therefore, the $d$-th delay tap of the delay domain MIMO channel matrix  ${\tilde{\mathbf{H}}_i[d]}$ can be written as
\begin{align}
\begin{array}{*{20}{l}}
%\hspace{-2mm}
{{\tilde{\mathbf{H}}_i}}\left[ d \right] =&{\beta_{0}^i{(d)}{\mathbf{a}}_r^i\left( {\theta _{0}^{r},\phi _{0}^{r}} \right){\mathbf{a}}_t^i{{\left( {\theta _{0}^{t},\phi _{0}^{t}} \right)}^H}}
\\
&+\sum\limits_{c = 1}^{{N_c}} {\sum\limits_{p = 1}^{{N_p}} {\beta _{c,p}^i{(d)}{\mathbf{a}}_r^i\left( {\theta _{c,p}^{r},\phi _{c,p}^{r}} \right){\mathbf{a}}_t^i{{\left( {\theta _{c,p}^{t},\phi _{c,p}^{t}} \right)}^H}} },
\end{array}
\end{align}
where subscript $i=\{1,2\}$ corresponds to the BS-RIS channel or RIS-UE channel, respectively. In other words,  ${\tilde{\mathbf{H}}_1[d]}$ and ${\tilde{\mathbf{H}}_2[d]}$ are two different samples generated from the same model in (1) independently. $\beta_{0}^i{(d)}=\sqrt { \frac{{{{N_T}/{N_R}}}}{ {{L}}   }}{\alpha_{0}^ip\left(dT_s\right)}$ represents the channel coefficient of LoS component and $\beta_{c,p}^i{(d)} = \sqrt { \frac{{{{N_T}{N_R}}}}{ {{L}}   }}{\alpha_{c,p}^i} p\left(  dT_s - \tau_{c,p}^i \right)$ is the delay-domain channel coefficient of the ${p}$-th path in the ${c}$-th scattering cluster. $L={N_c}{N_p}+1$ stands for number of the total rays in this channel model. $N_T$ and $N_R$ are the numbers of transmit antennas and receive antennas, respectively, and that gives us ${\tilde{\mathbf{H}}_1[d]} \in {\mathbb{C}}^{N_u \times N_t}$ and ${\tilde{\mathbf{H}}_2[d]} \in {\mathbb{C}}^{N_r \times N_u}$ in the downlink transmission. Additionally, $\alpha _{0}^i \sim {\mathop{{\cal C}{\cal N}}\nolimits} \left(0,1\right)$, $\alpha _{c,p}^i \sim {\mathop{{\cal C}{\cal N}}\nolimits} \left(0,10^{- {\mu}}\right)$ are the complex gains, where ${\mu}$ is the power distribution ratio of LoS to NLoS multipath components.  $p(\tau)$ is the pulse shaping filter for $T_s$ spaced signaling and $\tau_{c,p}^i$ is the relative time delay.
${\theta _{c,p}^{r}\left( {\phi _{c,p}^{r}} \right)}$ and ${\theta _{c,p}^{t}\left( {\phi _{c,p}^{t}} \right)}$ are the azimuth (elevation) AoAs and AoDs of the ${p}$-th path in the ${c}$-th scattering cluster, respectively.
The angles in each cluster follow the uniform distribution and have the constant angle spreads (standard deviation),
which can be  denoted by $\sigma _\phi ^{t}$, $\sigma _\theta ^{t}$, ${\sigma _\phi ^{r}}$, and $\sigma _\theta ^{r}$, respectively.
${{\mathbf{a}}_r^i\left( {\theta _{c,p}^{r},\phi _{c,p}^{r}} \right)}$ and ${{\mathbf{a}}_t^i\left( {\theta _{c,p}^{t},\phi _{c,p}^{t}} \right)}$ are the normalized receive and transmit array response vectors.
In the case of a  UPA in the ${yz}$-plane with ${N_y}$ and ${N_z}$ elements on the ${y}$ and ${z}$ axes, respectively, the array response vector can be written as
\begin{align}
\label{equ:model1}
\!\!\!\!\!{{{\mathbf{a}}_{{\mathrm{UPA}}}}\!\!\left( {\theta ,\phi } \right)} \!&= \!\frac{1}{{\sqrt {{N_y}{N_z}} }}\big[1, \cdots \!,{e^{j\frac{{2\pi }}{\lambda }d\left( {n\sin \left( \theta   \right)\!\cos \left( \phi \right) + m\sin \left( \phi  \right)} \right)}},\notag\\
&\cdots ,{e^{j\frac{{2\pi }}{\lambda }d\left( {\left( {N_y - 1} \right)\sin \left( \theta \right)\cos \left( \phi   \right) + \left( {N_z - 1} \right)\sin \left( \phi   \right)} \right)}}\big]^T,
\end{align}
where ${1 \le n < N_y}$ and ${1 \le m < N_z}$ are the ${y}$ and ${z}$ indices of an antenna element, respectively. Besides, ${\lambda }$ and ${d = \frac{\lambda}{2}}$ denote the wavelength and adjacent antenna spacing, respectively.

In the frequency domain, the channel response at the $k$-th subcarrier can be further written as
\begin{align}
{{\mathbf H }_i\left[ k \right] }= \sum\limits_{d = 0}^{D_l - 1} {{\tilde {\mathbf H}}_i\left[ d \right]{e^{ - j\frac{{2\pi k}}{K}d}}},
\end{align}
where $D_l$ is the length of the cyclic prefix (CP).

During the transmission phase, data symbols at the transmitter ${\mathbf{x} }\left[ k \right]\in {{\mathbb{C}}^{ {N_s}\times{1} }}$ of ${k}$-th subcarrier, $k = \{1, \cdots, K\}$, are first beamformed through the baseband beamformer ${{{\mathbf F}_{\mathrm {BB}}}[k]} \in {{\mathbb{C}}^{ {M_t}\times{N_s}}}$, and then the symbol blocks are transformed to the time domain by $K$-point inverse fast Fourier transform (IFFT) for each RF chain. After adding a CP, analog beamformer ${{\mathbf F}_{\mathrm {RF}}}\in {{\mathbb{C}}^{ {N_t}\times{M_t} }}$ is applied to form the transmit signal. Then the signal travels through the mmWave multipath channel  ${{\mathbf H }_1\left[ k \right] }$ to reach the RIS. Each element at the RIS can induce an adjustable phase ${\phi_{u}, u=\{1,\cdots,N_u\}}$ to the signals received independently. Here, we assume that the RIS can only adjust the phase for the incident signals. Accordingly, the effect of RIS can be modeled as a diagonal matrix  ${\mathbf \Phi} =$ diag${\left(\left[ {e^{j\phi_{1}} \, \cdots\,e^{j\phi_{N_u}}} \right]\right) }$. At the receiver, analog combiner ${{\mathbf W}_{\mathrm {RF}}\in{{\mathbb{C}}^{N_r \times M_r}}}$ is first used to receive the signal reflected by RIS, when the CP is further removed followed by $K$-point FFT, then the frequency domain signal at   $k$-th subcarrier can be combined using baseband combiner ${{{\mathbf W}_{\mathrm {BB}}}[k]}$. Therefore, in the downlink transmission, the received signal at the UE associated with the $k$-th subcarrier can be expressed as
\begin{equation}
\label{equ:test}
{{\mathbf y}\left[k\right]\!\!= \!\!{{\mathbf W}}_{\mathrm {BB}}^H\!\left[ k \right]} {{\mathbf W}_{\mathrm {RF}}^H} {\left( {{\mathbf H}_2\!\left[ k \right]}{\mathbf \Phi} { {\mathbf H}_1{\left[k\right]}} {{\mathbf F}_{\mathrm {RF}}\!{{\mathbf F}}_{\mathrm {BB} }\!\left[ k \right]\!{\mathbf{x}}\!\left[ k \right]\!\!+\!\!{{\mathbf{n}}}\left[ k \right]} \right)},
\end{equation}
where
${{\mathbf{n}}[k]}\in {{\mathbb{C}}^{{N_r} \times 1}}$ is the additive white Gaussian noise (AWGN) vector at the UE.

\section{Proposed Scheme}
In this section, we first put forward the design scheme for digital baseband beamformer and combiner at the UE/BS simultaneously, then we propose an oversampling 2D-DFT codebook based  analog beamformer/combiner design, where the SOMP algorithm is utilized to search for multiple optimal analog beams from the codebook. Finally, the idea of reflecting coefficient matrix  design at the RIS is presented. For the first two parts, the perfect CSI is assumed to be known at both the UE and BS. To be specific, the effect of the first stage channel ${{{\mathbf H}_1} \left[k\right]} $, phase shift matrix ${\mathbf \Phi }$, and the second stage channel ${ {{\mathbf H}_2} \left[k\right]}$ are equivalently denoted by ${{{\mathbf{H}}_{\mathrm {eff}} \left[k\right]}={{{{\mathbf H}_2} \left[k\right]}}{\mathbf \Phi}{{{\mathbf H}_1} \left[k\right]}}$, with ${{\mathbf H}_{\mathrm {eff}}[k]}\in {{\mathbb{C}}^{ {N_r}\times{N_t}}}$. As for the RIS, acquiring accurate NLoS information of channels is not easy.
%and  configuring the meta-surface in a quick response time is not easy.
Besides, for the mmWave frequency band in the future mobile communication, BSs and RISs are expected to be deployed densely. Therefore, the signal at the receiver is probably to contain a strong LoS path for both BS-RIS link and RIS-UE link. Thus, we will only use AoA and AoD of LoS component to design the RIS reflecting matrix.

\subsection{GMD-Based Baseband Beamformer/Combiner Design}
\label{sec:gmd}
For fully-digital MIMO systems, the SVD-based channel beamforming with waterfilling algorithm has been proved to be the optimal scheme to achieve the maximum  SE. However, when it comes to BER performance, it should be awared that waterfilling will further aggravate the gain difference among multiple parallel spatial domain subchannels. If the same modulation and coding schemes are adopted on different spatial-domain subchannels, BER performance is mainly determined by the subchannel with the lowest SNR. Hence, to reach a relatively low BER, subtle modulation and coding scheme need to be done in different subchannels, which will increase the design burden of both the transmitter and receiver.

For hybrid MIMO over frequency-selective-fading channels, although digital baseband beamformer and combiner on different subcarriers can be designed independently, the analog beamformer and combiner still need to be formulated jointly.
%Still, we can not avoid the complicated bit allocation caused by different SNR conditions among different spatial-domain subchannels.
It has been shown in \cite{gmd} that, for narrowband situation, by applying the GMD to the equivalent baseband channel, cooperating with the successive interference cancellation (SIC), we can achieve a better BER performance than the conventional SVD with waterfilling, in the case that no complicated adaptive modulation and coding schemes are designed for both of them.
Inspired by this idea, we expand the GMD-based beamforming scheme to the wideband case with the help of RIS, so that both the good BER and high SE can be achieved.

The implementation of GMD is based on SVD, and we use ${ {\mathbf {\hat{H}}  }[k] }$ to indicate the equivalent baseband channel at the ${k}$-th subcarrier, where ${{\mathbf {\hat H}} \left[k\right]}  = {{\mathbf W}_{\mathrm {RF}}^H} {{{ \mathbf H}_{\mathrm  {eff}}}{\left[k\right]}} {{\mathbf F}_{\mathrm {RF}}}$. Since the baseband beamformings can be designed independently for different subcarriers, we omit the subcarrier index $k$ of  ${{\mathbf {\hat H}} \left[k\right]}$, and then an SVD on channel matrix $\mathbf {\hat H}$ can be denoted by
\begin{equation}
\label{equ:svd}
{{\mathbf {\hat H}} }={{\mathbf {U} {\mathbf{\Sigma} }}{{\mathbf V}^H}}=
{\left[\begin{array}{*{20}{l}}\!
	{\mathbf U}_1&{\mathbf U}_2\\
	\end{array}
	\right]}
{\left[\begin{array}{*{20}{l}}\!
	{\mathbf {\Sigma}}_1&0\\
	0&{\mathbf {\Sigma}}_2
	\!\!\end{array}
	\right]}
{\left[\begin{array}{*{20}{l}}
	{{\mathbf V}_1}^H\\
	{{\mathbf V}_2}^H
	\end{array}
	\!\right]}.
\end{equation}
Moreover, when we iteratively adjust the diagonal element in ${\mathbf \Sigma}$ through permutations and Givens transformations \cite{GMDref}, this process can be realized through unitary matrices ${\mathbf S}_{L}$ and ${\mathbf S}_{R}$, expressed by
\begin{subequations}
	\begin{equation}
	{{\mathbf Q}_1}={\mathbf V}_1{{\mathbf S}_R},
	\end{equation}
	\begin{equation}
	{\mathbf G}_1={{\mathbf U}_1}{{\mathbf S}_L},
	\end{equation}
	\begin{equation}
	{\mathbf R}_1={{\mathbf S}_L^T{\mathbf \Sigma}_1}{{\mathbf S}_R}.
	\end{equation}
\end{subequations}
Accordingly, we can decompose the channel matrix in the GMD form, then the equivalent baseband channel at each subcarrier can be decomposed as
%{\color{red} the form of SVD should be present for alogrithm initialize}
\begin{equation}
\label{equ:gmd}
{{\mathbf {\hat H}} }={{\mathbf {GR}}{{\mathbf Q}^H}}=
{\left[\begin{array}{*{20}{l}}\!
	{\mathbf G}_1&{\mathbf G}_2\\
	\end{array}
	\right]}
{\left[\begin{array}{*{20}{l}}
	{\mathbf R}_1& {\mathbf R}_3\\
	0&{\mathbf R}_2
	\end{array}
	\right]}
{\left[\begin{array}{*{20}{l}}
	{{\mathbf Q}_1}^H\\
	{{\mathbf Q}_2}^H
	\end{array}
	\!\!\right]},
\end{equation}
where ${{{\mathbf G}_1} \in{\mathbb{C}^{M_r \times N_s}}}$ and
${{{\mathbf Q}_1} \in{\mathbb{C}^{M_t \times N_s}}}$ are semi-unitary matrices containing the left ${N_s}$ columns of ${{\mathbf G} \in {\mathbb{C}^{M_r \times M_r}}}$ and ${{\mathbf Q} \in {\mathbb{C}^{M_t \times M_t}}}$, respectively. ${{{\mathbf R}_1} \in {\mathbb{C}^{N_s \times N_s}}}$ is an upper triangular matrix whose diagonal elements are identical and equal to the geometric mean of the largest ${N_s}$ singular values of ${\mathbf R}$, i.e. $r_{i,i}=\bar{r}=({{\sigma_1}{\sigma_2}\cdots{\sigma_{N_s}}})^{\frac{1}{N_s}}$ holds for all the diagonal elements $r_{i,i}$ in ${{\mathbf R}_1}$, where $1\!\leq i\!\leq N_s$. ${{\mathbf R}_2}$ and ${{\mathbf R}_3}$ are irrelevant matrices and  they are not utilized for data transmission. Employing ${{\mathbf{F}}_{\mathrm {BB}}[k]}\!\!=\!\!{{\mathbf Q}_1[k]}$ as the baseband beamformer and ${{\mathbf W}_{\mathrm {BB}}^H[k]}\!=\!{{\mathbf G}_1^H[k]}$ as the baseband combiner for each subcarrier, then $(\ref{equ:test})$ can be rewritten as
\begin{align}
\begin{array}{*{20}{l}}
\label{equ:digitalp}
{{\mathbf y}[k]}&={\mathbf G}_1^H[k]{(\hat{\mathbf H}[k]{{\mathbf Q}_1[k]}{{\mathbf x}[k]}+{{\mathbf W}_{\mathrm {RF}}^H}{{\mathbf n}[k]})}
\\
\\&={{\mathbf R}_1[k]{{\mathbf x}[k]}}+{\mathbf G}_1^H[k]{{\mathbf W}_{\mathrm {RF}}^H}{{\mathbf n}[k]}.
\end{array}
\end{align}

Since the equivalent channel after baseband beamforming is not a diagonal matrix, SIC is necessary  at the receiver to cancel the interference among different data streams. Till now, the digital part design of hybrid MIMO structure has been finished and bit allocation can be effectively avoided.

\subsection{Oversampling Codebook Based SOMP Analog Beamforming}
\label{sec:codebook}

Since the digital baseband design has been completed in Section \ref{sec:gmd}, in this subsection, we mainly focus on the design of analog beamformer/combiner. An OMP based spatially sparse beamforming was proposed for the narrowband MIMO system in \cite{sparse}, hence, we consider a extended version of OMP in wideband situation, namely, SOMP. In conventional  OMP-based hybrid beamforming design, the main target is to choose $N_s$ optimal  beams from $L$ steering vectors associated with $L$ multipath components. In other words, the top ${N_s}$ subchannels with the largest singular values are chosen, then the baseband beamforming can be designed through a least squares way to approach the performance of fully-digital structure. However, the sensing matrix in this spatially sparse beamforming scheme requires the steering vectors of both NLoS and LoS paths, which are not easy to acquire as mentioned before.
%Besides, a GMD-based baseband beamforming scheme has a priority under practical condition
Thus, we put forward a mixed SOMP and GMD based hybrid beamforming scheme, in which an oversampling codebook is used.
Note that we mainly focus on the design of analog beamformer at the transmitter, and the analog combiner at the receiver can be acquired in a similar way.

\subsubsection{Problem Formulation}
As it is proved in Lemma${\;}$1 of \cite{gmd}, a sensing matrix ${{\mathbf A}_t=\left[{{\mathbf a}_t \left(\theta _{1}^{t}, \phi _{1}^{t}\right)},{{\mathbf a}_t \left(\theta _{2}^{t}, \phi _{2}^{t}\right)},\cdots,{{\mathbf a}_t \left(\theta _{L}^{t},\phi_{L}^{t}\right)}\right]}$ whose columns are composed of all the steering vectors at the BS side, is able to span the column of the unconstrained fully digital beamformer matrix at each subcarrier, denoted by ${{{\mathbf F}_{\mathrm {opt}}}\left[k\right],k=\{1,2,\cdots,K\}}$ and ${{{\mathbf F}_{\mathrm {opt}}}[k]}$ is the first $N_s$ columns of the right singular matrix of ${\mathbf H}_{\mathrm {eff}}[k]$. Since the ${{\mathbf A}_t}$ has satisfied the constant modulus constraint, we only need to choose $M_t$ optimal beams from ${{\mathbf A}_t}$ to form the analog beamformer ${\mathbf F}_{\mathrm {RF}}$. Therefore, we can use a selection matrix ${\mathbf T} \in {\mathbb{C}^{L\times{M_{t}}}}$ to form this procedure, i.e., ${{\mathbf F}_{\mathrm {RF}}={\mathbf A}_t{\mathbf T}}$. By exploiting the common angular-domain sparsity of mmWave massive MIMO channels over different subcarriers \cite{gao}, the analog beamforming problem can be formulated as follows
%\begin{equation}
\begin{align}
\label{equ:analog}
&{\mathbf T}=\mathop {\arg\min }\limits_{\mathbf{T}} \sum\limits_{k = 1}^K {\| {{\mathbf F}_{\mathrm {opt}}\left[k\right]}-{{{\mathbf A}_t{\mathbf T}}{\mathbf F}_{\mathrm {BB}}\left[k\right]}\|}_F \notag,\\
&{\mathrm{s}}{\mathrm{.t}}{\mathrm{. }}\:{\|\; } {\text {diag}}{\left({{\mathbf T}{\mathbf T}^H}\right) \|}_0{\mathrm{ = }}{ {M_t} }{\mathrm{,}}
%\quad {\left({\mathbf A}_t\right)_{i,j}=\frac{1}{\sqrt{N_t}} ,\forall i,j},
\end{align}
%\end{equation}
where ${{\mathbf A}_t}$ naturally satisfies the constant modulus constraint, since it is formed by all the steering vectors at the BS side. However, it is actually impossible acquire such ${{\mathbf A}_t}$ in practice. Therefore, we use a codebook scheme instead.
Besides, ${M_t}$ and $M_r$ are usually assumed to be smaller than $L$, so $\mathbf T$ will be a sparse matrix and sparse recovery algorithm such as SOMP can be used.

\subsubsection{Oversampling Codebook Design}
Due to the limited resolution of conventional DFT codebook, an oversampling codebook is considered for more refined spatial resolution \cite{codebook}. For the UPA, we use the ${\rho} $ to represent the oversampling factor, then the distinguishable angles divide the space into ${\rho N_y }$ discrete elevation angles in the vertical direction and ${\rho N_z}$ discrete azimuth angles in the horizonal direction. A notation description to this process can be expressed as follows.

We use ${\mathcal{R}_y}$ and ${\mathcal{R}_z}$ to represent the phases set that lie on the grid, which is given by ${\mathcal{R}_y=\{0,\frac{2\pi}{\rho N_y},\cdots,\frac{2\pi\left(\rho N_y-1\right)}{\rho N_y}\}}$ and ${\mathcal{R}_z=\{0,\frac{2\pi}{\rho N_z},\cdots,\frac{2\pi\left(\rho N_z-1\right)}{\rho N_z}\}}$, then all the combinations from  ${\mathcal{R}_y}$ and ${\mathcal{R}_z}$ form the candidates set of an oversampling codebook ${\mathbf D}$, i.e.,
\begin{align}
\label{codebook}
{\mathbf D}=\{{\mathbf{a}}_{{\mathrm{UPA}}}\left(\theta,\phi\right)|\theta\in\mathcal{R}_y,\phi \in \mathcal{R}_z,\forall \theta,\phi\}.
\end{align}

In this way, both the BS and UE can generate a codebook according to their respective oversampling factors. We assume that the oversampling factors are identical for both BS and UE in this article. Subsequently, we may use ${{\mathbf D}_t}$ to substitute ${{\mathbf A}_t}$ in $\left(\ref{equ:analog}\right)$ and adjust the size of selection matrix ${\mathbf T}$ accordingly. The codebook ${{\mathbf D}_r}$ at the receiver can be acquired in the same way. With the increasement of the oversampling factor $\rho$, the codebook has more candidates with more refined spatial resolution, so the quantization error between the true angles and their nearest candidates can be reduced. In this way, the matching bases with smaller quantization error to true angles can be searched.
According to the angular-domain common sparsity of different subcarriers \cite{gao}, when we fix ${{\mathbf F}_{\mathrm {BB}}\left[k\right]}$, we can treat the oversampling codebook as the basis and find multiple optimal beams to match the  ${\mathbf F}_{\mathrm {opt}}\left[k\right]$, and ${{{\mathbf F}_{\mathrm {opt}}}[k]}$ is the first $N_s$ columns of the right singular matrix of ${\mathbf H}_{\mathrm {eff}}[k]$. To sum up, the design of analog beamformer can be summarized in $\mathbf {Algorithm} \: \mathbf {\ref{alg:alg2}}$ and the analog combiner ${{\mathbf W}_{\mathrm {RF}}}$ at the receiver can be acquired in the same way.
\begin{algorithm}[t]
	\caption{Proposed Analog Beamforming Design}
	\label{alg:alg2}
	\begin{algorithmic}[1]
		\renewcommand{\algorithmicrequire}{\textbf{Input:}}
		\renewcommand\algorithmicensure {\textbf{Output:} }%{\mathrm{Empty Matrix}}
		\REQUIRE ~~\\
		Optimal beamformer ${{\mathbf F}_{\mathrm {opt}} \left[k\right]}, 1<k<K$, the number of RF chains $M_{t}$, and the oversampling factor $\rho$.
		\ENSURE ~~ \\
		Analog Beamformer ${{\mathbf F}_{\mathrm {RF}}}$.	
		\STATE Generate  codebook ${\mathbf D}_t$  according to (\ref{codebook});
		\label{step:1}
		\STATE Initialize the residual matrix ${\mathbf F}_{\mathrm {res}}[k]={\mathbf F}_{\mathrm {opt}}[k], \;\forall k$,  the index set ${\mathcal{A}_t}=\emptyset$ and analog beamformer ${\mathbf F}_{\mathrm {RF}}=\emptyset$;
		\label{step:2}
		\STATE \textbf{for} $i\underline{\;}iter = 1:M_{t}$
		\label{step:3}
		\STATE ~~~${\mathbf \Psi[k]}={\mathbf D}_t^H{{\mathbf F}_{\mathrm   {res}}[k]}{{\mathbf F}_{\mathrm {res}}^H[k]}{\mathbf D}_t, \forall k$;
		\label{step:4}
		\STATE ~~~$\:i=\mathop{\arg\max}\limits_{l=1,\cdots,L} {\left({\sum\limits_{k=1}^K{\mathbf \Psi[k]}}\right)}_{l,l}$;
		\label{step:5}	
		\STATE ~~~${\mathcal{A}_t=\mathcal{A}_t\cup {i},}$ and ${ {{\mathbf F}_{\mathrm {RF}}}=\{{\mathbf F}_{\mathrm {RF}},{\mathbf D}_t(:,i)\}};$
		\label{step:6}
		\STATE ~~~${\mathbf Y}[k]={{\mathbf F}_{\mathrm {RF}}}^\dagger {{\mathbf F}_{\mathrm {opt}}[k]}, \forall k$;
		\label{step:7}
		\STATE ~~~${{\mathbf F}_{\mathrm {res}}}[k]=\frac{{{\mathbf F}_{\mathrm {opt}}[k]}-{\mathbf F}_{\mathrm {RF}}{{\mathbf Y}[k]}}{\|{{\mathbf F}_{\mathrm {opt}}[k]}-{\mathbf F}_{\mathrm {RF}}{{\mathbf Y}[k]}\|_F}, \forall k$;
		\label{step:8}
		\STATE \textbf {end for};
		\label{step:9}
	\end{algorithmic}
\end{algorithm}
\subsection{Reflection Matrix Design for RIS}
\label{sec:ris}
RIS is composed of a large number of passive reflecting elements.
This leads to the prohibitive channel training overhead, and it is impractical to estimate the full-dimensional channels. In some cases, a quantized codebook based reflecting matrix can be designed without explicit channel estimation. However, the size of codebook still becomes very large owing to the large number of passive elements at the RIS.
In order to simplify the model and avoid the dilemma that we need to know the complete CSI when we design the passive beamforming for RIS, we only utilize the AoA of LoS BS-RIS link and AoD of LoS RIS-UE link seen from the RIS to design the reflecting coefficient matrix $\mathbf \Phi$. Here the associated AoD and AoA estimation can be obtained by state-of-the-art solutions \cite{{add2},{add3},{add4},{add5}}.

Let us assume that the AoA from the LoS BS-RIS link and AoD from the LoS RIS-UE link can be both well estimated. When we neglect the NLoS components and assume the AoAs and AoDs keep unchanged during the data transmission phase, then the LoS component of equivalent channel between the BS and UE link via RIS , denoted by ${\mathbf H}_{\mathrm {eff}}^{\mathrm {Los}}[k]$, can be expressed as
\vspace{-5mm}
\begin{align}
{{\mathbf H}_{\mathrm {eff}}^{\mathrm {Los}}[k]}={\zeta_2[k]\zeta_1[k]}{{\boldmath{\mathbf a}}_{r}^2(\theta_{0}^r,\phi_{0}^r)}\gamma{{\boldmath{\mathbf a}}_{t}^1(\theta_{0}^t,\phi_{0}^t)}^H,
\end{align}
where
\begin{align}
\gamma={{\boldmath{\mathbf a}}_{t}^2(\theta_{0}^t,\phi_{0}^t)}^H{\mathbf \Phi}{{\boldmath{\mathbf a}}_{r}^1(\theta_{0}^r,\phi_{0}^r)},
\\{\zeta_1[k]}= \sum\limits_{d = 0}^{D_l - 1} {\beta_{0}^1[d]{e^{ - j\frac{{2\pi k}}{K}d}}},
\\ {\zeta_2[k]}= \sum\limits_{d = 0}^{D_l - 1} {\beta_{0}^2[d]{e^{ - j\frac{{2\pi k}}{K}d}}}.
\end{align}
For convenience, we omit the angles $\phi$ and $\theta$ in steering vectors ${{\boldmath{\mathbf a}}_{t}^i(\theta_{0}^t,\phi_{0}^t)}^H$ and ${{\boldmath{\mathbf a}}_{r}^i(\theta_{0}^r,\phi_{0}^r)}$ in the following expression. Here the superscripts ``$1$" and ``$2$" for variables ${\mathbf a}_{r}^i$ and ${\mathbf a}_{t}^i$ correspond to the steering vector in BS-RIS link ${{\mathbf H}_1[k]}$ and the RIS-UE link ${{\mathbf H}_2[k]}$, respectively. The subscript ``$0$" means these parameters correspond to the LoS path component  and the $\gamma$ here can be equivalently seen as the array gain for the LoS component. Since the subchannels share the common angular-domain sparsity, the $\gamma$ can be the same for all subcarriers. Therefore, our optimization goal is to maximize the modulus of $\gamma$ with the constraint that the modulus of diagonal elements of ${\mathbf \Phi}$ are limited to one, which can be formulated as
\begin{equation}
\begin{aligned}
\label{equ:phase}
&{\mathbf \Phi}=\mathop {\arg\max }\limits_{\mathbf\Phi} \vert{\gamma}\vert,\\
&{\mathrm{s}}{\mathrm{.t}}{\mathrm{. }} \quad \vert {\left({\mathbf\Phi}\right)_{j,j}\vert=1 ,\forall j}.
\end{aligned}
\end{equation}

Since $\gamma$ is a scalar, and the steering vectors are normalized, thus, the ${\gamma}$ can be written as
\begin{equation}
\label{term}
{{\gamma}}=\sum\limits_{j = 1}^{N_u}{{(\boldmath{{\mathbf a}_t^2})_j}^{*}{\left({\mathbf\Phi}\right)_{j,j}}{\boldmath({\mathbf a}_r^1})_j}.
\end{equation}
Here each term of the product in (\ref{term}) is a scalar, whose absolute value is a constant. Therefore, it is quite easy to get the solution to $(\ref{equ:phase})$, which is given by
${\left({\mathbf\Phi}\right)_{j,j}}={{(\boldmath{\mathbf a}_t^2)_j}}{(\boldmath{\mathbf a}_r^1)_j^*}, \forall j$.
for all subcarriers.

\section{Simulation Results}
\label{sec:results}

In this section, we will compare the BER and SE performance of our proposed scheme with existing schemes in aspects of baseband beamforming, analog beamforming, and RIS reflection matrix design.

In simulations, the BS is equipped with a ${N_t=8 \times 8} $ UPA, the UE is equipped with a ${N_r =4\times 4}$ UPA, and the RIS is an $N_u=16 \times 16$ passive array. The carrier frequency is set to 28 GHz, and the antenna spacing is the half of carrier frequency wavelength. The number of subcarriers is $K=64$ and the CP length is $ D_l = 64$. For the wideband mmWave channels, the parameters for two stage channels are both set as follows: in addition to one LoS path, the number of NLoS clusters is assumed to be $N_c=7$ and each cluster has $N_p=10$ propagation paths, azimuth/elevation AoAs and AoDs follow the uniform distribution  ${\cal U}\left[ { - \pi /2,\pi /2} \right]$, the angle spreads are $\sigma _\phi ^{t}=\sigma _\theta ^{t}= {\sigma _\phi ^{r}} =\sigma _\theta ^{r}=7.5^{\circ}$. The paths delay is uniformly distributed in ${\cal U}\left[ { 0, D_lT_s} \right]$. In mmWave system, channel power of LoS path is much higher than NLoS paths, therefore, similar to that in \cite{gao}, the power distribution ratio $\mu$ is set to $2.3$ in the simulations, which means the power of LoS component is 23 dB higher than that of NLoS components. Besides, we consider the 16-QAM modulation in all simulations, the number of data streams, RF chains at the transmitter and receiver are all set to the same value, i.e., $N_s = M_{t} = M_{r} = 3$.

\begin{figure}
	\centering
	\includegraphics[width=8cm, keepaspectratio]%
	{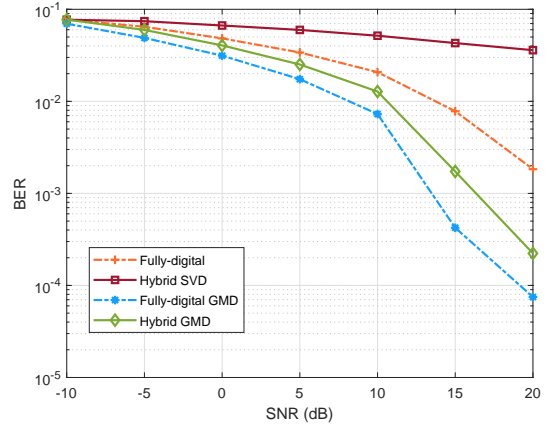}
	\vspace*{-1.0mm}
	\caption{BER performance of two different baseband beamforming schemes for both hybrid MIMO and fully-digital MIMO structures with the assistance of RIS. In hybrid MIMO structure, both of the two schemes share the same SOMP-based analog beamforming using all steering vectors as a prior information.}
	\label{fig2}
	\vspace*{-3.0mm}
\end{figure}

Fig. \ref{fig2} compares two different baseband beamforming schemes for the scenario as shown in Fig. \ref{fig1}. SVD baseband beamforming in  Section \ref{sec:gmd} is compared as benchmark. For fully-digital MIMO structure, the analog beamforming is unnecessary. For hybrid beamforming at both the BS and UE, both ``ideal SOMP based analog beamforming" is adopted for both GMD and SVD  baseband beamforming schemes. Here, ``ideal SOMP based analog beamforming" means the sensing matrix ${\mathbf A}_t$ is composed of all the real steering vectors, which can be hard to perfectly acquire in practice. Besides, the reflection coefficient matrix of RIS is designed in Section \ref{sec:ris}.

It can be seen from Fig. \ref{fig2} that the GMD baseband scheme outperforms the SVD scheme in both hybrid and fully-digital MIMO structure, because GMD-based beamforming can alleviate the SNR variation among different spatial-domain  sub-channels, then the overall BER performance deterioration due to some low SNR channels can be avoided, compared with SVD-based schemes without carefully designed bit allocation.

Besides, the gap between the fully-digital structure MIMO and hybrid structure MIMO is smaller for GMD-based scheme.

\begin{figure}
	\centering
	\includegraphics[width=8cm, keepaspectratio]%
	{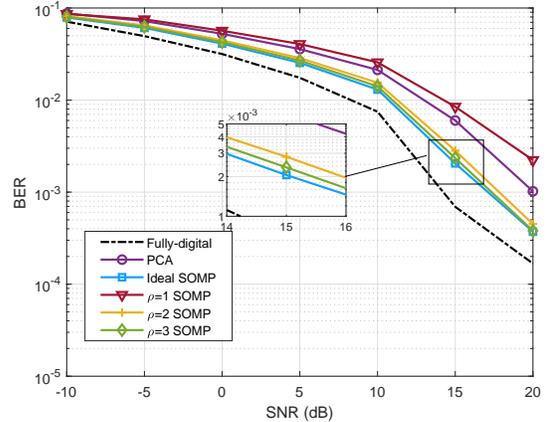}
	\vspace*{-1.0mm}
	\caption{BER performance of different analog beamforming schemes in wideband MIMO system with the assistance of RIS, where the same GMD-based baseband beamforming is applied to all the schemes above.}
	\label{fig3}
	\vspace*{-3.0mm}
\end{figure}

Fig. \ref{fig3} compares the BER performance achieved by different analog beamforming schemes for the scenario in Fig.\ref{fig1}. Without specially indicated, we will only consider GMD-based baseband beamformer and combiner in the following figures due to its advantage over SVD.
In Fig. \ref{fig3}, three different schemes using the same GMD baseband beamforming in Section \ref{sec:gmd} are compared as benchmarks. They are:
1) {\textit {Fully digital}}: fully-digital MIMO structure without analog beamforming;
2) {\textit{Ideal SOMP}}: hybrid MIMO structure with ideal SOMP based analog beamforming using all steering vectors as a prior information;
3) {\textit {PCA}}: hybrid MIMO structure with principal component analysis (PCA)-based analog beamformer in \cite{pca}.
Our proposed codebook-based analog beamforming scheme in Section \ref{sec:codebook} is distinguished by different oversampling factors $\rho$, which is marked as ``$\rho=r$ SOMP", $r=\{1,2,3\}$ in the figure. All the schemes above adopt the proposed  RIS reflection matirx in Section \ref{sec:ris}.

\begin{figure}
	\centering
	\includegraphics[width=8cm, keepaspectratio]%
	{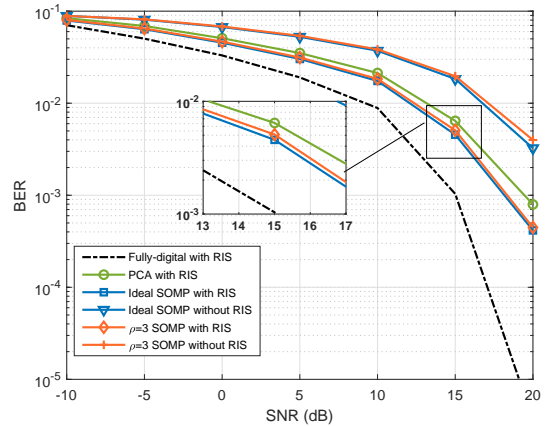}
	\vspace*{-1.0mm}
	\caption{BER performance comparison between Fig. \ref{fig1} scenario using RIS ( $\mu = 2.3$ ) and NLoS scenario without using RIS, and the same GMD baseband beamforming is applied to all the schemes above.}
	\label{fig4}
	\vspace*{-3.0mm}
\end{figure}
We can observe from Fig. \ref{fig3} that, when the oversampling factor ${\rho}$ is greater than 2, the proposed oversampling codebook based SOMP analog beamforming scheme outperforms the PCA scheme. Meanwhile, when the codebook with a larger $\rho$ is adopted, its BER performance approaches the {\textit {Ideal SOMP}} scheme. When the $\rho $ is greater than 3, the proposed scheme performs nearly as well as the {\textit {Ideal SOMP}} scheme.

Different from Fig. \ref{fig1} scenario, Fig. \ref{fig4} also discusses a NLoS scenario, which shows the performance improvement brought by the RIS. We compare the scenario in Fig.\ref{fig1} with the NLoS scenario that a direct multipath channel link is established between the BS and UE without the help of RIS. However, the LoS path is blocked by the buildings in the later scenario. GMD baseband beamforming is applied to all the schemes, too.
The results show that by designing of RIS reflecting matrix, we can achieve the better BER performance than the NLoS environment without the assistance of RIS.

\begin{figure}
	\centering
	\includegraphics[width=8cm, keepaspectratio]%
	{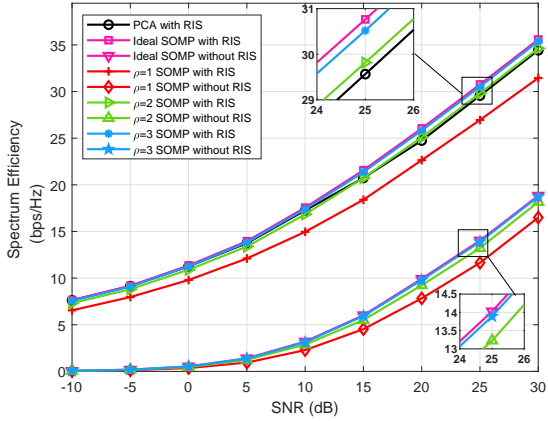}
	\vspace*{-1.0mm}
	\caption{SE of different analog beamforming schemes in Fig. \ref{fig1} scenario ( $\mu = 2.3$ ) and NLoS scenario, which respectively correspond to the occasion with and without the assistance of RIS. The same GMD beamforming is applied in the baseband.}
	\label{fig5}
	\vspace*{-5.0mm}
\end{figure}

Fig. \ref{fig5} illustrates the SE performance of the proposed schemes against different SNRs. Similar to Fig. \ref{fig4}, Fig. \ref{fig1} scenario and NLoS scenario are both compared under different analog beamforming schemes, and the same GMD baseband beamformer/combiner is applied to all the schemes.
As is shown in Fig. \ref{fig5}, the proposed scheme ``$\rho=2$ SOMP with RIS" and ``$\rho=3$ SOMP with RIS" outperforms ``PCA with RIS" at high SNR. In addition, the proposed scheme can perform nearly as well as the ideal SOMP scheme when $\rho$ is greater than 3. Besides, Fig. \ref{fig5} also shows that the assistance of RIS can improve the SE at the same time. This is because when the RIS is utilized, the  LoS component can be effectively exploited, a virtual LoS link between the BS and UE via RIS can be established for improved channel quality and transmission performance. However, when the RIS is not considered, only NLoS components can be exploited between the BS and UE, and the channel quality is poor with degraded BER performance.  % which give us the insights that more delicated RIS design can futher improve the performance.

\begin{figure}
	\centering
	\includegraphics[width=8cm, keepaspectratio]%
	{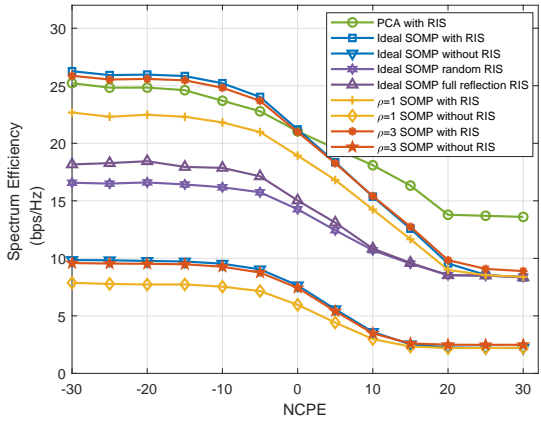}
	\vspace*{-1.0mm}
	\caption{SE performance with perturbation, where the SNR = $20$ dB and the same GMD baseband beamforming is applied.}
	\label{fig6}
	\vspace*{-3.0mm}
\end{figure}

Fig. \ref{fig6} compares the robustness of different schemes with imperfect CSI. We consider different analog beamforming schemes in Fig. \ref{fig1} scenario and NLoS scenario with same GMD baseband beamforming in this figure. The normalized channel perturbation error (NCPE) is used to model the channel perturbation, which includes channel estimation error, the CSI quantization in channel feedback, and/or outdated CSI. The NCPE is defined as
\begin{equation}
\begin{aligned}
{NCPE}=\frac{{\sum\limits_{k = 1}^K {\| {{\mathbf H}_i[k]}-{{\mathbf H}_{i}^e[k]}\|_F^{2}}}}{{\sum\limits_{k = 1}^K {\| {{\mathbf H}_i[k]}\|_F^{2}}}}, i=1,2,
\end{aligned}
\end{equation}
where $i=1$ and $i=2$ stand for the BS-RIS link channel and RIS-UE link channel, respectively. ${{\mathbf H}_{i}^e[k]}={{{\mathbf H}_i[k]} + {{\mathbf N}_i[k]}}$, ${{\mathbf N}_i[k]}$ is the perturbation noise vector, and each element of $ {{\mathbf N}_i[k]}$  follows the complex Gaussian distribution ${\mathop{{\cal C}{\cal N}}\nolimits} \left(0,\sigma_{e}^2\right)$. As it is shown in Fig. \ref{fig6}, the proposed scheme shows some robustness to SE variation with the increasement of the perturbation error. When we set the SNR to 20 dB, the SE of ``$\rho\!=\!3$ SOMP with RIS" stays at a high stable value when the NCPE is less than 0 dB, which shows the robustness to the channel perturbation error.
In addition, in Fig. \ref{fig6}, we also evaluate the SE performance of different RIS design schemes. Specifically, the legend ``Ideal SOMP full reflection RIS"  means we apply the fully reflection matrix to the incident signal, i.e., ${\mathbf \Phi} = {\mathbf I}_{N_u}$. ``Ideal SOMP random RIS" represents that we give each  diagonal element of $\mathbf \Phi$ a random phase following uniform distribution ${\cal U}\left[ {0, 2\pi} \right]$. Simulation results enlighten us that with more delicate reflecting matrix design, we can achieve a better SE performance.

\section{Conclusions}
This paper has proposed a hybrid beamforming scheme for wideband mmWave MIMO systems with the assistance of RIS.  To be specific, for the  baseband part, by utilizing GMD, we propose a baseband beamforming design for each subcarrier, where we can get a better BER performance without complicated bit allocation as that in the traditional SVD-based schemes. For the  analog part, by using the SOMP analog beamforming based on an oversampling 2D-DFT codebook, we acquire the analog beamforming design at both the BS and UE, which solves the difficulties of wideband MIMO beamforming and the prerequisites to obtain all the steering vectors in the traditional sparse beamforming schemes. Furthermore, we give an RIS reflecting matrix design scheme according to the AoA and AoD associated with the LoS BS-RIS channel and RIS-UE channel, which can harvest the large array gain for improved BER and SE performance.
%, which also shows robust performance to the error of channel estimation.
Finally, simulation results show that the proposed scheme can achieve a good performance in wideband hybrid MIMO system with the assistance of RIS.


\begin{thebibliography}{00}
\bibitem{summary}
E. Basar, M. Di Renzo, J. De Rosny, M. Debbah, M.-S. Alouini, and R. Zhang, ``Wireless communications through reconfigurable intelligent surfaces," \textit{IEEE Access}, vol. 7, pp. 116753-116773, Aug. 2019.
\bibitem{liao}A. Liao, Z. Gao, H. Wang, S. Chen, M. -S. Alouini and H. Yin, ``Closed-loop sparse channel estimation for wideband millimeter-wave full-dimensional MIMO systems," \textit{ IEEE Trans. Commun.}, vol. 67, no. 12, pp. 8329-8345, Dec. 2019.

\bibitem{gao_cs}Z. Gao, L. Dai, S. Han, C.-L. I, Z. Wang, and L. Hanzo, ``Compressive sensing techniques for next-generation wireless communications," \textit{IEEE Wireless Commun.}, vol. 25, no. 4, pp. 144-153, Jun. 2018. 	
\bibitem{add1}L. Wan, L. Sun, X. Kong, Y. Yuan, K. Sun, F. Xia, ``Task-driven resource assignment in mobile-edge computing exploiting evolutionary computation," \textit {IEEE  Wireless Commun.}, vol. 26, no.6, pp. 94-101, Dec. 2019.

\bibitem{irs_beamforming}Q. Wu and R. Zhang, ``Intelligent reflecting surface enhanced wireless network via joint active and passive beamforming," \textit{IEEE Trans. Wireless Commun.}, vol. 18, no. 11, pp. 5394-5409, Nov. 2019.

\bibitem{irs1}Q. Wu and R. Zhang, ``Joint active and passive beamforming
optimization for intelligent reflecting surface assisted SWIPT under QoS constraints," [Online]. Available: https://arxiv.org/abs/1910.06220.

\bibitem{irs2}H. Guo, Y. Liang, J. Chen and E. G. Larsson, ``Weighted sum-rate optimization for intelligent reflecting surface enhanced wireless networks," [Online]. Available: https://arxiv.org/abs/1905.07920.

\bibitem{irs3}P. Wang, X. Yuan, Z. Chen, H. Duan and H. Li, ``Intelligent reflecting surface-assisted millimeter wave communications: Joint active and passive precoding design," [Online]. Available: https://arxiv.org/abs/ 1908.10734.
\bibitem{irs4}W. Yan, X. Kuai, and Xiao. Yuan, ``Passive Beamforming and Information Transfer via Large Intelligent Surface," [Online]. Available: https://arxiv.org/abs/1905.01491.
\bibitem{ke}M. Ke, Z. Gao, Y. Wu, X. Gao, and R. Schober, ``Compressive sensing based adaptive active user detection and channel estimation: Massive access meets massive MIMO," to appear in \textit{IEEE Trans. Signal Process.}
\bibitem{irs_ofdm}B. Zheng and R. Zhang, ``Intelligent reflecting surface-enhanced OFDM: Channel estimation and reflection optimization," \textit{IEEE Wireless Commun. Lett.}, vol. PP, no. PP, Dec. 2019. doi: 10.1109/LWC.2019.2961357.

\bibitem{wan}Z. Wan, Z. Gao, B. Shim, K. Yang, G. Mao and M. -S. Alouini, "Compressive sensing based channel estimation for millimeter-wave full-dimensional MIMO with lens-array," \textit {IEEE Trans. Veh. Technol.}, doi: 10.1109/TVT.2019.2962242.
\bibitem{gmd}T. Xie, L. Dai, X. Gao, M. Z. Shakir and J. Li, ``Geometric mean decomposition based hybrid precoding for millimeter-wave massive MIMO," \textit{China Commun.}, vol. 15, no. 5, pp. 229-238, May. 2018.
\bibitem{sparse}O. E. Ayach, S. Rajagopal, S. Abu-Surra, Z. Pi and R. W. Heath, ``Spatially sparse precoding in millimeter wave MIMO systems," \textit{IEEE Trans. Wireless Commun.}, vol. 13, no. 3, pp. 1499-1513, Mar. 2014.
\bibitem{somp}C. H. Chen, C. Tsai, Y. Liu, W. Hung, and A. Wu, ``Compressive sensing (CS) assisted low-complexity beamspace hybrid precoding for millimeter-wave MIMO systems," \textit {IEEE Trans. Signal Process.}, vol. 65, no. 6, pp. 1412-1424, Mar. 2017.

\bibitem{codebook}J. Mao, Z. Gao, Y. Wu and M. -S. Alouini, ``Over-sampling codebook-based hybrid minimum sum-mean-square-error precoding for millimeter-wave 3D-MIMO," \textit{IEEE Wireless Commun. Lett.}, vol. 7, no. 6, pp. 938-941, Dec. 2018.
\bibitem{GMDref}C.-E. Chen, Y.-C. Tsai, and C.-H. Yang, ``An iterative
geometric mean decomposition algorithm for MIMO communications systems," \textit{IEEE Tran. Wireless Commun.}, vol. 14, no. 1, pp.343–352, Jan. 2015.
\bibitem{greedy}J. A. Tropp, A. C. Gilbert, and M. J. Strauss, ``Algorithms for simultaneous sparse approximation: Part I: Greedy pursuit," \textit{Signal Process}. vol. 86, pp. 572–588, Mar. 2006.
\bibitem{pca}Y. Sun, Z. Gao, H. Wang, B. Shim, G. Gui, G. Mao and F. Adachi, ``Principal component analysis based broadband hybrid precoding for millimeter-wave massive MIMO systems," [Online], Available: https://arxiv.org/abs/1903.09074.
\bibitem{heath1}K. Venugopal, N. González-Prelcic and R. W. Heath, ``Optimal frequency-flat precoding for frequency-selective millimeter wave channels," \textit{Trans. Wireless Commun.}, vol. 18, no. 11, pp. 5098-5112, Nov. 2019.
\bibitem{heath2}A. Alkhateeb and R. W. Heath Jr., ``Frequency selective hybrid precoding for limited feedback millimeter wave systems," \textit{IEEE Trans. Commun.}, vol. 64, no. 5, pp. 1801-1818, May. 2016.
\bibitem{heath4}S. Park, A. Alkhateeb, and R. W. Heath, ``Dynamic subarrays for hybrid precoding in wideband mmWave MIMO system," \textit{IEEE Trans. Wireless Commun.}, vol. 16, no. 5, pp 2907-2920, May. 2017.
\bibitem{dai}J. Tan, L. Dai, J. Li and S. Jin, ``Angle-based codebook for low-resolution hybrid precoding in millimeter-wave massive MIMO systems," \textit{IEEE/CIC International Conference on Communications (ICCC)}, Qingdao, China, pp. 1-5, Oct. 2017.

\bibitem{estimation1}Z. He and X. Yuan, ``Cascaded channel estimation for large intelligent metasurface assisted massive MIMO," \textit{IEEE Wireless Commun. Lett.}, vol. PP, no. PP, Dec. 2019. doi: 10.1109/LWC.2019.2948632.
\bibitem{heath3}J. Rodríguez-Fernández, N. González-Prelcic, K. Venugopal, and R. W. Heath, Jr., ``Frequency-domain compressive channel estimation for frequency-selective hybrid millimeter wave MIMO systems," \textit{IEEE Trans. Wireless Commun.}, vol. 17, no. 5, pp. 2946–2960, May. 2018.
\bibitem{fei}B. Wang, F. Gao, S. Jin, H. Lin, and G. Y. Li, ``Spatial- and frequency wideband effects in millimeter-wave massive MIMO systems," \textit {IEEE Trans. Signal Process.}, vol. 66, no. 13, pp. 3393–3406, Jul. 2018.
\bibitem{gao}Z. Gao, C. Hu, L. Dai and Z. Wang, ``Channel estimation for millimeter-wave massive MIMO with hybrid precoding over frequency-selective fading channels," {\textit {IEEE Commun. Lett.}} vol. 20, no. 6, pp. 1259-1262, June. 2016.

\bibitem{add2}H. Wang, L. Wan, M. Dong, K. Ota, and X. Wang, ``Assistant vehicle localization based on three collaborative base stations via SBL based robust DOA estimation," \textit {IEEE Internet  Things J.}, vol. 6, no. 3, pp. 5766-5777, June. 2019.
\bibitem{add3}X. Wang, L. Wan, M. Huang, C. Shen, and K. Zhang. ``Polarization channel estimation for circular and non-circular signals in massive MIMO systems," \textit{IEEE J. Sel. Topics  Signal Process.}, vol. 13, no. 5, pp. 1001-1016, Sept. 2019.
\bibitem{add4}L. Wan, X. Kong, and F. Xia. ``Joint range-Doppler-angle estimation for intelligent tracking of moving aerial targets", \textit{IEEE Internet Things J.}, vol. 5, no.3, pp 1625-1636, June. 2018.
\bibitem{add5}F. Wen, J. Shi, Z. Zhang. ``Joint 2D-DOD, 2D-DOA and polarization angles estimation for bistatic EMVS-MIMO radar via PARAFAC analysis," \textit{IEEE Trans. Veh. Technol.}, doi: 10.1109/TVT.2019.2957511.	
\end{thebibliography}
\end{document}